# Security Assessment Rating Framework for Enterprises using MITRE ATT&CK® Matrix

Hardik Manocha[1] · Akash Srivastava[2] · Chetan Verma[2] · Ratan Gupta[2] · Bhavya Bansal[2] ·

**Abstract**
Threats targeting cyberspace are becoming more prominent and intelligent day by day; this inherently leads to a dire demand for continuous security validation and testing. Using this paper, we aim to provide a holistic and precise security analysis rating framework for organizations that increases the overall coherency of the outcomes of such testing. This scorecard is based on the security assessment performed following the globally accessible knowledge base of adversary tactics and techniques called the MITRE ATT&CK matrix. The scorecard for an evaluation is generated by ingesting the security testing results into our framework, which provides an organization's overall risk assessment rating and the risk related to each of the different tactics from the ATT&CK matrix.

## 1 Introduction

Information technologies have become a critical sector of post-industrial society in all realms of human affairs, increasing information streams in industries and organizations' external and internal environment. The increase in the number of enterprises and their climbing rates foreordains the need for cybersecurity systems in there is a dire need for security assessment and administration. A prime principle of security assurance is the implementation of strategic planning for risk reduction.

Hardik Manocha
hardik@fourcore.vision

Akash Srivastava
akashs1698@gmail.com

Chetan Verma
chetanverma99.cv@gmail.com

Ratan Gupta
itsgupta.ratan@gmail.com

Bhavya Bansal
bhavya.bansal1998@gmail.com

[1] FourCore Labs, New Delhi

[2] Bharati Vidyapeeth's College of Engineering, New Delhi

Excluding the indomitable risks that arise from the supply chain or the production processes, the other fundamental solution to risk reduction is to have continuous security validation and testing of the organizations. In addition, security Teams must protect critical infrastructure powering their organizations. To curb this demand, organizations generally invest in Breach and Attack simulation security solutions and extensive organizational security vulnerability testing. These provide a comprehensive, detailed analysis of the corporate landscape regarding the security risks and vulnerabilities and the remediation of these security flaws.

The continuous security testing and reports, however detailed they are, are only focused on the Security Operations Centres teams (SOC) and security analysts. However, these tactics come up short on the methods for evaluating the elements of administration and those modelling tactics to prevent information security breaches. The reports make much too little sense to the administrative units. The other consequence of these rising breaches is that organizations are having trouble relying on other businesses. That is where security ratings come to play. The organization receives a personalized scorecard that provides a quantitative analysis of security posture in the industry, prioritizing the indicators and production dynamics that need the required attention. These security ratings are independently correlatable with the breach risk and stock performance. The ratings offer extensive visibility into the critical areas of cyber risk associated with the breach, including Compromised Systems, Open Ports, Mobile & Desktop Software, and File Sharing. Thus, providing a more extensive view into the attack surface can be prioritized and collaborated internally to address the most significant areas of cyber risk.

Security ratings can be used to assist risk management teams in supervising cybersecurity risk, together with:

- Comprehension of third and fourth-party risk posed by supply chain, third-party vendor, and business partner relationships.
- Cyber insurance underwriting, pricing, and risk management allow insurers to gain visibility into the security program of those they insure to assess better and price their insurance policies.
- Investment in or acquire a company by providing organizations with an independent assessment of an investment or M & M&A target's information security controls.
- Enabling governments to understand better and manage their vendors' cybersecurity performance is critical in

- FISMA compliance.
- Aid in supplementing and sometimes replacing time-consuming vendor risk assessment techniques like questionnaires, on-site visits, and penetration tests. Most importantly, they are always up to date.

By providing cybersecurity teams with the knowledge to identify security issues instantly, they can understand which vendors to focus on first. This dramatically reduces the operational burden on Third-party risk management committees throughout vendor selection, integration, and management. In addition, they can be accorded with vendors to enhance remediation efforts.

Such a method of security assessment scoring is presented below. The paper is structured into the following segments: related work in risk assessment scoring, the principal problem statement, and methodology involved followed by the results and conclusions.

## 2 Related Work

Defining a security metric is one of the challenging fields in security. Multiple parameters are considered when determining a security metric.

Kerameti et al. (2013) proposed defining some security metrics by combining the CVSS framework and attack graph. The prime problem with the existing approaches is that they cannot take interrelation between vulnerabilities of the network effectively and efficiently. Their defined security metrics can provide this issue and help security analysts to assess network security quantitatively by analysing attack graphs. Based upon their security metrics analysing model, they were able to find the most dangerous vulnerability in the network [1].

Again Zhang, M. et al. (2016) has proposed a general method for modelling a security metric based on a network diversity by evaluating and then designing a series of diversity metrics. This proposal first devised the biodiversity-inspired metric based on the adequate number of different resources and the proposed two complementary diversity metrics. Based on the most petite, average attacking efforts. As a result, they devised an effective richness-based metric based on its counterpart [2].

Based on the CVSS Metrics, Umesh Kumar Singh and Chanchala Joshi (2016) proposed a model for risk estimation that uses the vulnerability database 'National Vulnerability Database (NVD)' and the 'Common Vulnerability Scoring System'. The proposed model provides quantitative security metrics that produce fast and consistent security assessment, which helps in automated and reasonable security management. This paper presents the improved Quantitative CVSS Risk Level Estimation Model, which effectively estimates vulnerability risk. The proposed model computes the overall risk level of a system based on maturity and frequency estimates. In addition, a new dimension is introduced for calculating the frequency of the vulnerability with the assumption that the more frequent occurrences of vulnerability make the system riskier [3].

Marcus Pendleton et al. (2017), concerns about how to determine the security at the system level by developing a security metric based on these four sub metrics: (1) metrics of system vulnerabilities, (2) metrics of defense power, (3) metrics of attack or threat severity, and (4) metrics of situations. To analyse the relationship between these four sub-metrics, they propose a hierarchical ontology with four sub-ontologies corresponding to the four sub-metrics and discuss how they relate [4].

Talking about web application security, Nichols, E. A., & Peterson, G. (2007) targeted the security of web applications. Over the years, web application's functionality and user base have improved, and the rise in their vulnerabilities. Even though network firewalls are present to protect web applications against various cyber-attacks, they are insufficient to provide overall Web application security. The proposed research has provided metrics that can help quantify the impact of process changes in one life-cycle phase on other phases. The application's life cycle was divided into three main stages: design, deployment, and runtime. By organizing metrics according to the life cycle and the OWASP type, they could get insights from the derived quantitative results that can potentially point to defective processes and suggest improvement strategies [5].

Considering the scorecard-based assessment in mobile phones, Shabe et al. (2017) have studied the current state of cyber-security awareness among cell phone users living in South Africa. They proposed a security scorecard model which suggests how vulnerable individuals are to cyber-attacks. Since cell phones are used for almost every other task, it becomes crucial that a cell phone user's level of competence in security space must be improved. Therefore, a scorecard approach presents the measurements of cyber-security awareness among cell phone users [6].

Alex Ramos et al. (2017) provided a survey on the Model-based Quantitative Network Security Metrics. They present a deep study of the present proposals. First, to discriminate the security metrics described in this survey from other types of security metrics, an overview of security metrics, in general, and their classifications is presented. They also provided a detailed review of the main existing model based quantitative NSMs, along with their advantages and disadvantages. Finally, this survey is concluded with an in-depth discussion on relevant characteristics of the surveyed proposals and open research issues of the topic [7].

Eli Weintraub, Yuval Cohen (2018) have focused on the new measure called "Network Availability Exposure", which has not been considered in standard security risk scoring models. Network Availability Exposure shows the structure of the software/hardware components and their characteristics of the web and the relation between these components. This contributes to achieving good/bad network availability. The framework proposed enables getting accurate risk measures, thus allowing the organization to make better risk management decisions, allocating risk management budgets to the risks [8].

R. Fatkieva, A. Krupina (2020) has developed an algorithm, namely enterprise balanced scorecard development, and implementation. It provides a comprehensive assessment of the system's security aspects. The main objective behind this research is to create a methodological framework for security control built on a scorecard that considers indicators and production dynamics. Moreover, they have created a generalized description of various models and methods to establish their relationships and correspondences using multiple metrics. The research results' originality consists in the modelling of the impact of information security indicators, and the methods proposed cannot wholly replace the enterprise security control system; nevertheless, they are somewhat of real practical value due to the visualization of the strategy and to the monitoring of its implementation [9].

Rainer Diesch et al. (2020) have developed a comprehensive model of relevant management success factors (MSF) for organizational information security. First, they have surveyed various factors that directly or indirectly affect the system's safety. Then they interviewed 19 industry experts to evaluate the relevance of these vectors in practice and explore interdependencies between them. Furthermore, in the last phase, a model was developed which shows that there is key- security-indicators, which directly impact the security status of an organization while other indicators are only indirectly connected [10].

A. Aigner and A. Khelil (2006) have proposed a security analysis framework, called Security Qualification Matrix (SQM) for Cyber-Physical Systems (CPCs) such as the Internet of things, Smart Factories or Smart Grid. The security frameworks that are being used today for evaluating the security of CPCs may generate inaccurate security scores for CPS, as they consider the typical CPS characteristics, like the communication of heterogeneous systems of the physical and cyber-space domain in an unpredictable manner. At the same time, the proposed framework is capable of analysing multiple attacks on a System-of-Systems level simultaneously [11].

Amankwah et al. (2020) have focused on the security aspect of a web application. The standard vulnerability scoring system (CVSS) generates a score based on the application's security, but it has been challenged in previous studies, leading to various vulnerability scoring metrics. Hence authors have proposed an automated framework for evaluating open-source Web scanner vulnerability severity using a Web vulnerability detection scanner called zed attack proxy to detect vulnerabilities in a damn vulnerable web application [12].

Elissa M. Redmiles et al. (2020) have studied how much the security advice is perceivable and prioritized by users and experts. For this, they first conducted a large-scale, user-driven measurement study to identify unique recommended behaviours containing online security and privacy advice documents. Second, they develop and validate measurement approaches for evaluating the quality -- comprehensibility, perceived actionability, and perceived efficacy -- of security advice. Third, they deploy these measurement approaches for assessing the unique pieces of security advice in a user study with users and professional security experts. The results suggest a crisis of advice prioritization. Most of the advice is perceived by most users to be at least somewhat actionable and somewhat comprehensible. However, both users and experts struggle to prioritize this advice [13].

A. Aigner and A. Khelil (2020) authors have addressed the gap in the current security metrics framework that is being used to evaluate the security of Cyber-Physical Systems (CPS), System of Systems. Other complex systems by benchmarking a carefully selected variety of existing security metrics in terms of their usability for CPS. Also, in the process of benchmarking these frameworks, they have pinpointed critical CPS challenges and qualitatively assessed the effectiveness of the existing metrics for CPS systems [14].

Eleni Philippou et al. (2020) contextualizes the security metric to increase adaptability for organizational contexts, domain, technical infrastructure, stakeholders, business process, etc. They have proposed Symbiosis, a methodology that defines a goal elicitation and refinement process mapping business objectives to security measurement goals via systematic templates that capture relevant context elements. Their analysis shows that lack of understanding led to various high-profile security attacks in multiple organizations [15].

Sean Kinser et al. (2020) worked in the security aspect of the increasing commercial-space environment. They present a method known as the Architecture Score Index to quantitatively evaluate the overall trust of space-based services related to the core cybersecurity principles of confidentiality, availability, and integrity. This method considers both qualitative and quantitative assessments generated through phases categorized as Compliance Assessment, Performance of day-to-day cybersecurity operations (cyber-hygiene), and Incident Response. The inputs of these phases are used to generate a quantitative metric that indicates an organization's ability to deliver data and services [16] securely.

Jamal N. Al-Karaki, et al. (2020) authors present GoSafe, a novel practical cybersecurity assessment framework based on the ISO 2700x standard requirements for the Information Security Management System (ISMS). It can be used for self-assessment and auditing/scoring tools by national cybersecurity authorities. In the GoSafe framework, a novel mathematical model was also designed and implemented for the scoring/rating tool, namely, the national cybersecurity index (aeNCI). The aeNCI employs multiple parameters to determine the maturity of existing cybersecurity programs at national organizations and generate classification and comparison reports [17]

Tupper, Melanie & Zincir-Heywood (2008) proposed a novel quantitative security metric, VEA-ability, which measures the desirability of different network configurations. An administrator can then use the VEA-ability scores of different configurations to configure a secure network. Based on their findings, the framework can be used to estimate the comparative desirability of a specific network configuration accurately. This information can then be used to explore possible alternate configurations and allows an administrator to select one among the given options. These tools are essential to network administrators as they strive to provide secure yet functional network configurations [18].

Priscoli, Francesco et al. (2020) defined a security metric allowing evaluators to assess the security level of CPSs, in a holistic, consistent, and repeatable way. To achieve this, A mathematical framework provided by the open-source security testing methodology manual (OSSTMM) is taken as the base for the new security metric since it allows to offer security indicators capturing, in a non-biased way, the security level of a system.

## 3 Problem Statement: Building a Metric System for Managing and Monitoring Information Security Systems

Combining different models and approaches in assessing industry performance enables the creation of an information security control system. However, to construct such a system, it is vital to solving the following problems:
1. Evolution of an architectural plan of the organization's functioning.
2. Formulation of a generalized representation of a variety of models and techniques to establish their relationships using various benchmarks.
3. Classification and determination of a system of indicators for monitoring the events of processes.
4. Formalization of principles for the resolution of problems of decision for security assurance.

The solution to these above tasks allows developing a security management system that encompasses the elements for monitoring and conjecturing the organization's functioning, thus enabling us to reflect on the influences of the processes when these events and indicators change.

The idea is to create a comprehensive framework that collects various techniques from the globally accepted MITRE ATT&CK[21] matrix, which is a well-defined list of all the publicly known adversary tactics and techniques. Then, these techniques are assigned a severity score based on the impact they create on an organization and the complexity posed to an attacker, followed by providing their success/failure states from the collected reports and tests into our model. The outcome will be an organization's overall risk assessment rating and the respective coverage of the MITRE ATT&CK matrix per assessment.

## 4 Methodology

The formulation to build an approach to the contemporary assessment of the influences of indicators changing in an organization under the influence of security testing in a destructive approach requires two steps.
- Creation of a list of targeted goals focused on the production environment for the scorecard development. This involves breaking the entire assessment process into various stages with their corresponding goals and objectives. The next part is to deconstruct the above goals into subgoals to achieve a standardized performance vector. Further, we invest in demarking each goal to an impact and exploitability level widely categorized into three levels: Low, Medium, and High, thus allowing us to map the targeted goals to a generalized indicator directly. The dynamics of this process helps to access numerically to get a global average of the quantitative goals. To achieve this metric, we have used the globally accessible adversary knowledge base: MITRE ATT&CK Matrix.[21]
- Formulation of a method to generate a numeric figure that ingests the techniques processed from the ATT&CK matrix with the success/failure state from the targeted enterprise, evaluated against the total number of tactics, techniques, and their impact and exploitability from our labelled ATT&CK matrix and provide a generalized risk scorecard.

## 4.1 MITRE ATT&CK Matrix: A collection of targeted goals for risk analysis

Modern network attacks can be executed in various ways, making it ineffective to form a distinct description for each attack technique and its type. The increase in attack techniques grew a significant demand for a unified database maintaining these techniques and descriptions.

MITRE ATT&CK was developed in 2013 due to MITRE's Fort Meade Experiment (FMX) [20], where researchers emulated both adversary and defender behaviour to enhance post-compromise discovery of threats through telemetry and behavioural analysis.

MITRE ATT&CK® [21] stands for MITRE Adversarial Tactics, Techniques, and Common Knowledge (ATT&CK). It is a curated information base and standard for cyber adversary behaviour, exhibiting the different aspects of an adversary's attack lifecycle and the platforms they are identified to target. The tactics and techniques abstraction in the model provides a universal registry of unique adversary actions recognized by both offensive and defensive cybersecurity. It also presents an appropriate level of categorization for adversary activity and specific ways of defending against it.

The behavioural model presented by ATT&CK contains the following core elements:
- **Tactics:** tactical adversary goals during an attack.
- **Techniques:** describing how adversaries achieve tactical goals.
- Documented adversary usage of techniques and other metadata

An attack's lifecycle can be broken down into multiple smaller goals know as tactics:
1. Initial Access: various entry vectors to gain an initial foothold with a network. Some techniques may include exploiting public-facing applications, adding hardware, replicating through removable media, spear-phishing attachment/link, spear-phishing via service, compromising supply chain, trusted relationships, etc.
2. Execution: techniques that result in adversary-controlled code execution on a local or remote system. Some techniques may include using Command-Line Interface, Control Panel items or dynamic data exchange, through API/GUI/PowerShell/module load, compiled HTML file, scheduled task, scripting, service execution.
3. Persistence: techniques to maintain access across system changes. Some techniques may include account manipulation, application shimming, bootkit, browser extensions, changing default file association, external remote services, hidden files, and directories, hooking, image file execution options.
4. Privilege Escalation: techniques to gain greater permissions on a system or network. Some techniques may include access token manipulation, application shimming, bypassing user application control, DLL injection, exploitation for privilege escalation, extra window memory injection, process injection.
5. Defense Evasion: techniques to avoid detection during execution. Some techniques may include binary padding, clear command history, code signing, DLL search order hijacking, deobfuscate/decode files/information, disabling security tools, indicator blocking/removal, file deletion.
6. Credential Access: techniques to steal account credentials. Some techniques may include credential dumping, credentials in files, input capture, network sniffing, password filter DLL, two-factor authentication interception.
7. Discovery: techniques to gain intel about the system and network. Some techniques may include system services/information, applications, query registry, network configuration, remote systems, system owner/users, network services/connections, processes, security software.
8. Lateral Movement: techniques to enter and control remote systems on a network. Some techniques may include third-party/application deployment software, remote services, Windows remote management/Admin shares, logon scripts.
9. Collection: techniques to gather information from target systems. Some techniques may include data from local system/removable media/network shared drive, input/audio/video/screen capture, data staged.
10. Exfiltration: techniques to steal data from your network. Some techniques may include data compressed/encrypted, exfiltration over another network medium/command and control channel/alternative protocol/physical medium, automated exfiltration, scheduled transfer, data transfer size.
11. Command and Control: techniques to communicate with systems under their control within a victim network. Some techniques may include data obfuscation/encoding, fallback channels, custom cryptographic/command & control protocol, multiband communication, commonly/uncommonly used port, standard application/non-application layer protocol, multilayer encryption, connection proxy, communication through removable media.

The above categorization of different tactics and techniques in their unified group allows us to demarcate their main goals and subgoals uniquely.

We identify two terms, *Impact*: The relative risk/damage that a successful technique would have to a user's environment, and *Exploitability*: the complexity faced by the attacker to execute a technique successfully. In a broader approach, we can classify the following acts into their respective Impact and Exploitability.

**Table 4.1.1:** Classification of various goals against their Impact and Exploitability

| Goal | Impact and Exploitability |
|---|---|
| To escalate from a local privilege to a higher privilege | High impact, allowing higher-level access to specific objects and artifacts can be highly lethal, but high exploitability since very difficult to execute |
| To execute malware discreetly | High impact, hidden executions can be brutal to defend against. Medium exploitability since various methods are heavily documented online. |
| Supply Chain Compromise | High impact, non-mitigable risk factor. High exploitability since it is extremely difficult to execute. |
| Archive Collected Data | Low impact, the non-lethal activity of compressing data. Low exploitability by using the standard OS tools |

Following the publicly known Common Vulnerability Scoring System [22], we can define a Base score, their impact, and exploitability to the techniques of the ATT&CK matrix, which allows us to systematically visualize the impact of the targeted techniques (sub-goal) by the attack vector, its complexity, scope, privilege requirement, user-interaction, and the Impact Metric distributed between Confidentiality Impact, Integrity Impact, and Availability Impact.

The above metrics can be mapped to primary markers of impact levels and their respective exploitability distributed between low, medium, and high.

**Table 4.1.2:** Metric classification with reasoning

| Technique/Sub-techniques | Impact and Exploitability | Reasoning |
|---|---|---|
| Audio Capture (T1123) [Collection] | High, Medium | Adversary recording audio from the pre-installed or post-installed audio interfaces present in an operating system can lead to the potential risk of Intellectual Property Theft |
| Native API (T1106) [Execution, Defense Evasion] | High, Medium | Successful execution of malicious binary using Native API like Create Process, OpenProcess, and WriteProcessMemory can potentially evade various defensive systems in place, leading to malware execution. |
| Local Account Discovery (T1087.001) [Discovery] | Low, Low | Getting local account information on a system using command-line executions like whoami and dir is a very generic task performed by various applications |
| Keylogging (T1056.001) [Credential Access] | High, High | Keylogging involves recording keystrokes; this can disclose potential usernames and passwords to various internal and external services, thus high-risk. |
| DLL Search Order Hijacking (T1574.001) [Privilege Escalation] | High, High | DLL Search Hijacking can potentially lead to a higher privilege since the targeted process might be running in a high-privilege mode, but the application directory can have lower privileges. |
| AppCert DLLs (T1564.009) [Privilege Escalation/ Persistence] | Medium, Medium | establishing persistence and/or elevating privileges by executing malicious content triggered by AppCert DLLs can be achieved only on older/ outdated systems since AppCert DLL associations have been discarded from the latest Windows 10 OS. |

The above table represents some of the labelling markers for protection scores for all the techniques covered in the ATT&CK matrix, which becomes a testament to the facts backed up by industrial experiences and knowledge from various fields. The markers, however, can be refined on further maturity of this framework.

### 4.2 Formalisation of targeted goals to a single entity

To build a rating system based on decomposed structural tactics into various techniques, we create a framework for measuring and evaluating different attack vectors for each level at a given instant. This involves formulating a method to generate a protection score based on simulated attack assessments on the targeted enterprise, evaluated against the total number of tactics and techniques. This equation requires the convolution of indicators, in our case, Impact and Exploitability, to provide rapid score generation at any instant.

The initial step for formalizing the equation is to have an additive convolution with consistent weight coefficients for distinct pointers, which allows us to evaluate a generalized performance indicator for each technique to find the system's Protection Score.

Weighing each technique is further done using three parameters, namely, *High*, *Medium,* and *Low*. Finally, we can assign values to these parameters by using a 0-10 scale.

**Table 4.2.1:** Success/Failure score categorized by severity level of attack technique

| Status  | High | Medium | Low |
|---------|------|--------|-----|
| Success | 9    | 5      | 1   |
| Failure | 9.5  | 5.5    | 1.5 |

Also, the following statement is valid for each Technique's Protection Score.

$$Protection\ Score \propto Exploitability$$
$$Protection\ Score \propto 1/Impact$$

(a) The increase in Exploitability of a technique is an increase in the Protection Score since this represents the effort required to bypass the defense and protection systems.
(b) The increase in the Impact of a technique causes a decrease in the Protection Score since this represents the risk and damage incurred by a technique to a user's environment.

Using these, we plot the following graph for each parameter [E and I] considering the extreme conditions of
- High Impact and Exploitability (approx. 50%)
- Medium Impact, Low Exploitability (approx. 75%)
- High Exploitability, Low Impact (approx. 100%)

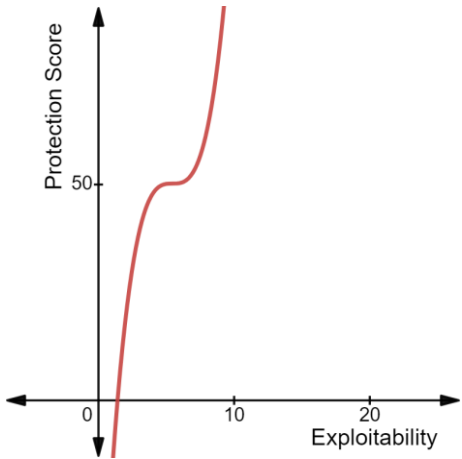

**Figure 4.1:** Exploitability vs Prediction Score

By evaluating this graph, we get the following equation,
$$f1 = ((E/a) - 5)^3 + 50 \quad \ldots (1)$$

Where,
E = Exploitability Weight
a = Graph Adjustment Constant = 1.1

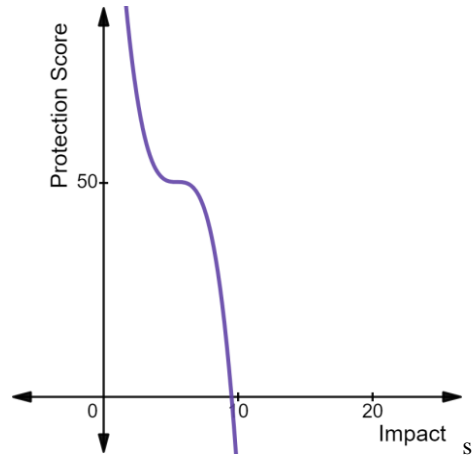

**Figure 4.2:** Impact vs Prediction Score

By evaluating this graph, we get the following equation,
$$f2 = -((I/a) - 5)^3 + 50 \quad \ldots (2)$$
Where,
I = Impact Weight
a = Graph Adjustment Constant = 1.1

We can further move to the convolution of our indicators, Impact, and Exploitability, to get the combined plot and the final equation.

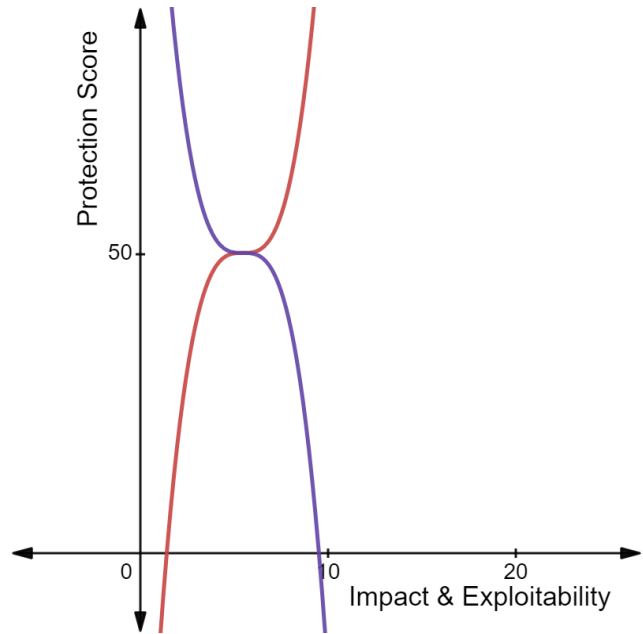

Figure 4.3: Impact and Exploitability vs Prediction Score

The above graph represents the combined curve of Exploitability (E) and Impact (I).

Using equation (1) and (2), we can now state the Protection Score as the mean of *f1* and *f2*

$$P = \frac{f1 + f2}{2} \quad ..(3)$$

therefore,

$$P = \frac{(\frac{E}{a}-5)^3 - (\frac{I}{a}-5)^3 + 100}{2} \quad ..(4)$$

Where,
P = Protection scores for one tactic
E = Exploitability Weight
I = Impact Weight
a = Graph Adjustment Constant = 1.1

Hence, we considered this formula along with (Tables 4.1.2 and 4.2.1) and created the table below. This table (Table 4.2.2) represents the calculated outcomes of the Exploitability and Impact Weights according to their success and failure.

**Table 4.2.2:** Success/Failure score categorized by severity levels of Exploitability and Impact

| Exploitability (E) | Impact (I) | Protection Score | |
|---|---|---|---|
| | | Success | Failure |
| High | High | 50% | 50% |
| High | Medium | 66% | 74% |
| High | Low | 100% | 98% |
| Medium | High | 34% | 26% |
| Medium | Medium | 50% | 50% |
| Medium | Low | 74% | 84% |
| Low | High | 0.33% | 2% |
| Low | Medium | 16% | 26% |
| Low | Low | 50% | 50% |

To get the effective *Protection Score for multiple techniques taken during an assessment*, we must use a **weighted arithmetic mean**. Therefore, we have divided the resultant Protection Score into five different categories known as *"Protection Categories"* and assigned weights in the following manner:

**Table 4.2.3:** Assigned weights to the 5 Protection Categories

| Protection Category | Very High | High | Medium | Low | Very Low |
|---|---|---|---|---|---|
| Weights Assigned | 1 | 0.8 | 0.5 | 0.2 | 0.1 |

Hence,

$$P_{total\ score} = \frac{\sum_{i=1}^{n} p_i * w_i}{\sum_{i=1}^{n} w_i}$$

Where,
W = Weight assigned to the Protection category.
$p_i$ = Protection Score of i'th tactic

Hence, using the above formulas, we can successfully calculate the Protection score for one technique or a group of techniques.

## 5 Results

Let us proceed with an example of an enterprise pen-testing assessment, the state space of which can be described in the following table (Table 5.1), where various techniques from different tactics of the MITRE ATT&CK Matrix have been considered:

1. We pick a case of a breach simulation. Hence, we start from Initial Access into the organization, followed by Client Execution on the target system using Native API utilizing Defense Evasion tactics such as Thread Local Storage: Process Injection techniques.
2. The attacker tried to maintain persistence on the target system using Application Shimming, which failed. The next attempt was to escalate privilege to the super admin using WinLogon Helper DLL, which failed again, as it was blocked by the Endpoint Detection and Response System (EDR).
3. The attacker successfully moved to a host in a different domain using Credential Stealing techniques like stealing Credentials from Registry followed by a network discovery and finally Lateral Movement through Remote Desktop Protocol, which concluded the assessment.

**Table 5.1:** Example Pentesting Assessment

| Technique Id | Technique Name | Exploitability | Impact | Success / Failure | Protection Score (%) |
|---|---|---|---|---|---|

| T1190 | Exploit Public-Facing Applications | High | High | S | 50 |
| T1106 | Native API | Med | High | S | 34 |
| T1055.005 | Thread Local Storage | Med | High | S | 34 |
| T1546.011 | Application Shimming | Med | Med | F | 50 |
| T1547.004 | WinLogon Helper DLL | Low | Med | F | 26 |
| T1552.002 | Credentials in Registry | Low | Med | S | 16 |
| T1135 | Network Share Discovery | Low | Low | S | 50 |
| T1021.001 | Remote Desktop Protocol | Med | High | S | 34 |

The above assessment provided the following Protection Score by ingesting the above results into the score calculator.

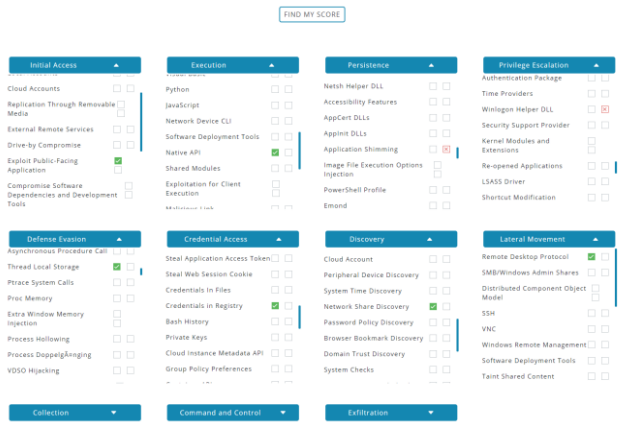

**Fig 5.1:** Score Calculator

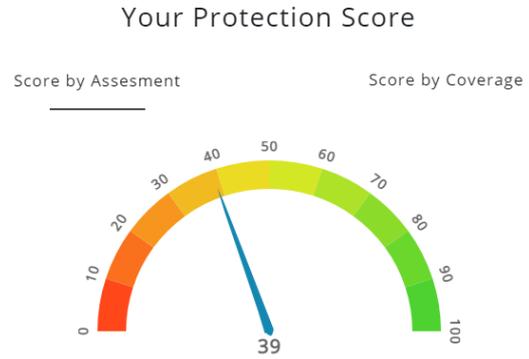

Security alright. But, put your security guys to work right now.

**Fig 5.2:** Protection Score: 39%

**Total Protection Score = 39 %**

**Score by Coverage = 1/100** (Coverage score represents the total coverage of the MITRE ATT&CK Matrix)

Like the above example, building a strategy plan aids in assessing the influence of every indicator on accomplishing the target goal, technique indicators convolution with their respective weight coefficients, damage evaluation, and construction of a set of actions capable of decreasing the risk. Damage caused by a particular technique is tricky to determine; multiple factors play a role in assessing the damage. Using the data points produced in our testing, we were able to plot a graph: Impact of a tactic vs Damage and Weighted Decreasing Protection vs Damage.

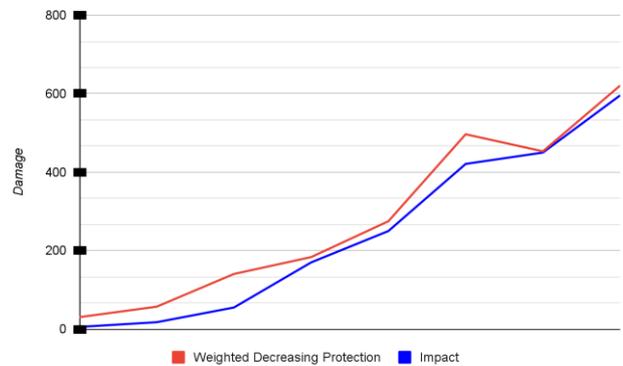

**Fig 5.3:** Combined Plot: Tactic Impact vs Damage vs Protection

The graph (Fig 5.3) shows a trend of the impact of a particular technique and its possible damage rating and the same trend, which is followed for the Protection Score for techniques. The most negligible impact on the ability of damage minimization has the indicator represented by the ratio of the number of techniques executed within an enterprise to the total number of the parts enterprise processes. To reduce the influence of such tactics and techniques on business performance, a set of measures can be considered. Increasing the number of security layers in the organization by at least 40% will significantly increase the exploitability of various techniques and attack vectors. Similarly, following a trusted and strategic supply chain following all the modern compliances will further help in reducing the total risk. Restrictions of employees' access to vital facilities and control of critical machinery operations further help strengthen the organization's security posture.

The additional efforts were undertaken to observe both uptime and downtime in the organization and reducing the risk of critical infrastructure breakdown, including that due to unapproved access, will successfully achieve a secure, trusted organization security landscape.

## 6 Conclusion and Future Works

In this research paper, we have developed an algorithm that determines the security protection score of an organization, providing an overall holistic view of the organizational security posture. The prominence of this score further helps increasing awareness and building strategic methods for information security assurance. It is further anticipated that this security rating framework will help in reducing the gap between the security operations team and the organization's strategic development team. The research results originality is based on industrial experience in the field of cybersecurity and operations; however, there is a vast space for improvement, which is why this project has been made open source for further enhancement, and available as a node NPM package for direct integration into other web platforms.